\documentclass[]{spie}  

\usepackage{amsmath,amsfonts,amssymb}
\usepackage{graphicx}
\usepackage[colorlinks=true, allcolors=blue]{hyperref}
\usepackage{amsmath,amsfonts,amssymb}
\usepackage{multicol}
\usepackage{tabu}
\usepackage{float}
\usepackage{textcomp,gensymb}

\title{Characterization of a silicon photomultiplier for the Ultra-Fast Astronomy telescope}

\author[a]{Siyang Li}
\author[a-g]{George F. Smoot}

\affil[a]{Department of Physics, University of California, Berkeley, USA}
\affil[b]{Lawrence Berkeley National Laboratory, USA}
\affil[c]{Department of Physics, Hong Kong University of Science and Technology, China}
\affil[d]{Institute for Advanced Study, Hong Kong University of Science and Technology, China}
\affil[e]{Energetic Cosmos Laboratory, Nazarbayev University, Kazakhstan}
\affil[f]{Department of Physics, Universit\'e Paris Diderot, France}
\affil[g]{Paris Centre for Cosmological Physics, Universit\'e Paris, France}

\authorinfo{Further author information: (Send correspondence to S.L.)\\S.L.: E-mail: seanli@berkeley.edu}

\begin{document} 
\maketitle

\begin{abstract}

We characterized the S13360-3050CS Multi-Pixel Photon Counter (MPPC), a silicon photomultiplier (SiPM) manufactured by Hamamatsu Photonics K.K.. Measurements were obtained inside a light tight dark box using 365 nm, 400 nm, 525 nm, 660 nm, 720 nm, 810 nm, and 900 nm light-emitting diodes (LED) and the Citiroc 1A front-end evaluation system manufactured by Weeroc. At a 2.95V over voltage, we measured a dark count rate of 5.07x10\textsuperscript{5} counts per second at 26\degree C, crosstalk probability of 8.7$\%$, photon detection efficiency of 36$\%$ at 400 nm, linear range of 1.8x10\textsuperscript{7} photons per second, and saturation at 5x10\textsuperscript{8} photons per second. The S13360-3050CS MPPC is a candidate detector for the Ultra-Fast Astronomy (UFA) telescope which will characterize the optical sky in the millisecond to nanosecond timescales using two SiPM arrays operated in coincidence mounted on the 0.7 meter Nazarbayev University Transient Telescope at the Assy-Turgen Astrophysical Observatory (NUTTelA-TAO) located near Almaty, Kazakhstan. One objective of the UFA telescope will be to search for optical counterparts to fast radio bursts (FRB) that can be used to identify the origins of FRB and probe the epoch of reionization and baryonic matter in the interstellar and intergalactic mediums.
\end{abstract}


\section{INTRODUCTION}

Silicon photomultipliers (SiPM) are p-n junction semiconductor photodetectors consisting of Geiger-mode single-photon avalanche photodiodes (SPAD) connected in parallel. SiPM are beginning to replace traditional photodetectors such as photomultiplier tubes (PMT) in fields that require fast, single-photon counting resolution such as medical imaging, microscopy, commercial sensing, particle physics, and astronomy. Compared to PMT, SiPM have similar gains, similar timing resolutions, higher photon detection efficiencies (PDE), lower operating biases, and lower cost per channel. These characteristics make SiPM an attractive candidate detector for pushing the limits of astronomy down to the nanosecond regime and searching for sub-second astrophysical transients. Previously, few measurements\cite{Horowitz2001, Eikenberry_1997, Leung:2018} in the sub-second time domain had been conducted largely due to the limiting read times and noise penalties of most array imagers such as charge coupled devices. SiPM and SPAD are now beginning to be used to search for sub-second astrophysical signals such as extraterrestrial technosignatures\cite{Li2019, Wright2018} and Cherenkov radiation.\cite{Rando2015}. 

The first generation Ultra-Fast Astronomy (UFA) telescope\cite{Li2019b} will characterize the optical sky in the millisecond to nanosecond timescales and search for optical counterparts to fast radio bursts (FRB) using two single-photon counting detectors operated in coincidence on the 0.7 meter Nazarbayev University Transient Telescope at the Assy-Turgen Astrophysical Observatory (NUTTelA-TAO)\cite{Grossan2019,ECL, ECLConf} located near Almaty, Kazakhstan. FRB are high energy millisecond duration radio transients of unknown origin. While several theories on the origins of FRB exist\cite{Geng2015,Kashiyama2013,Cordes2016}, none have been experimentally verified. Optical counterparts to FRB could be used to investigate the emission mechanisms and origins of FRB and would likely exist in the sub-millisecond timescale due to dispersion and pulse broadening. As FRB have high dispersion measures that point to an extragalactic origin, optical counterparts to FRB could also be used to probe the epoch of reionization and baryonic matter in the interstellar and intergalactic mediums. 

In this paper, we present a characterization of the dark count rate, crosstalk probability, PDE, linearity, and saturation of a new S13360-3050CS Multi-Pixel Photon Counter (MPPC), a SiPM manufactured by Hamamatsu Photonics K.K.. This characterization was performed in an effort to evaluate the feasibility and advantages of using the S13360-3050CS MPPC for the UFA telescope and astrophysical observations in general.

\section{EXPERIMENTAL SETUP}

\begin{figure} [H]
\begin{center}
\begin{multicols}{2}
\includegraphics[height=6cm]{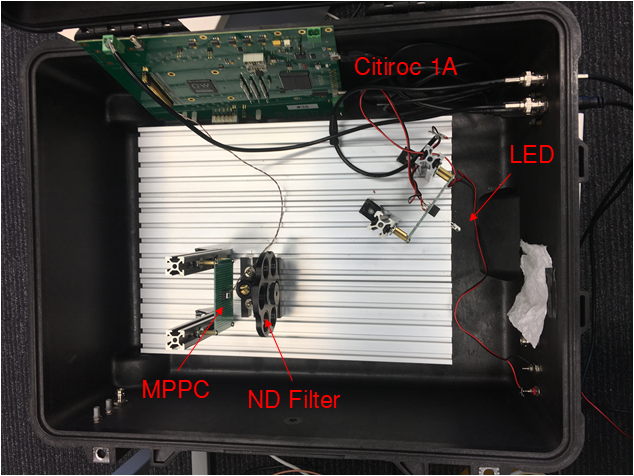}\par 
\includegraphics[height=6cm]{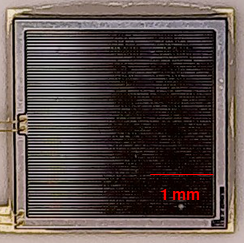}\par 
\end{multicols}
\end{center}
\caption{LEFT: Dark box test stand used to characterize the S13360-3050CS MPPC. RIGHT: S13360-3050CS MPPC from Hamamatsu Photonics K.K..}
\label{fig:darkbox}
\end{figure}

The experimental setup and the S13360-3050CS MPPC characterized in this study can be seen in Fig. \ref{fig:darkbox}. Detector parameters provided by Hamamatsu\cite{datasheet} can be seen in Table \ref{tab:parameters}. The MPPC was placed inside a light tight dark box with BNC, USB, and power supply feedthroughs and illuminated using scattered light from 365 nm, 400 nm, 525 nm, 660 nm, 720 nm, 810 nm, and 900 nm light-emitting diodes (LED). An ultraviolet enhanced silicon photodiode with a spectral range of 200 nm - 1100 nm was connected to a USB power meter and computer and used to obtain the photon flux at the location of the MPPC. The power meter has a 10 pW resolution and is calibrated with an uncertainty of 1$\%$ from 350 nm - 949 nm. The intensities of the LED were varied using a potentiometer and by adjusting the voltage of the LED. A neutral density (ND) filter with an optical density of 2.0 and an experimentally verified transmittance of 2$\%$ at 660 nm was used to obtain measurements with incident power below the 10 pW resolution of the power meter. 



\begin{table}[H]
\begin{center}
\begin{tabu}  { | X[c] | X[c] | }

\hline
\textbf{Parameter} &
\textbf{S13360-3050CS MPPC} \\
\hline
Spectral Range & 270 nm - 900 nm \\
\hline
Photosensitive Area  & 3.0 mm x 3.0 mm \\
\hline
Number of Pixels & 3600 \\
\hline
Pixel Pitch  & 50 $\mu$m \\
\hline
Fill Factor & 74$\%$ \\
\hline
Breakdown Voltage & 52.55V \\
\hline
Peak Photon Detection Efficiency & 40$\%$ at 450nm \\
\hline
\end{tabu}
\end{center}
\caption{Detector parameters of the S13360-3050CS MPPC characterized in this study.}
\label{tab:parameters}
\end{table} 

\begin{figure}[H]
\centering
\includegraphics[width=.9\textwidth]{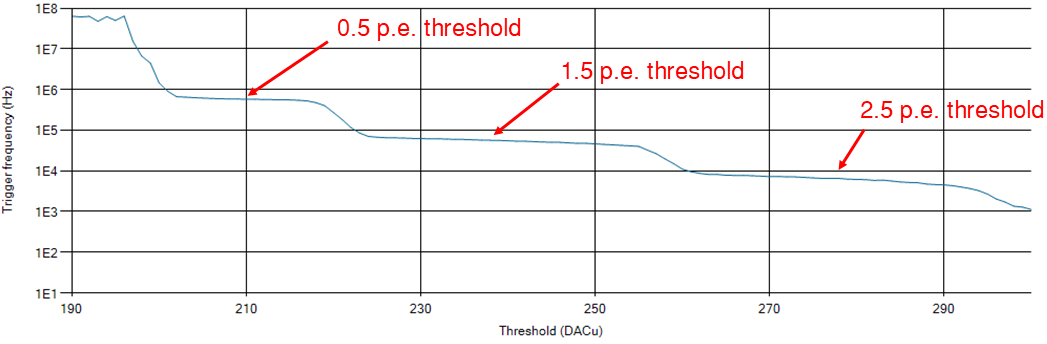}
\caption{Typical staircase plot obtained with the S13360-3050CS MPPC using the Citiroc 1A evaluation system.}
\label{fig:Staircase}
\end{figure}
\begin{figure}[H]
\centering
\includegraphics[width=.6\textwidth]{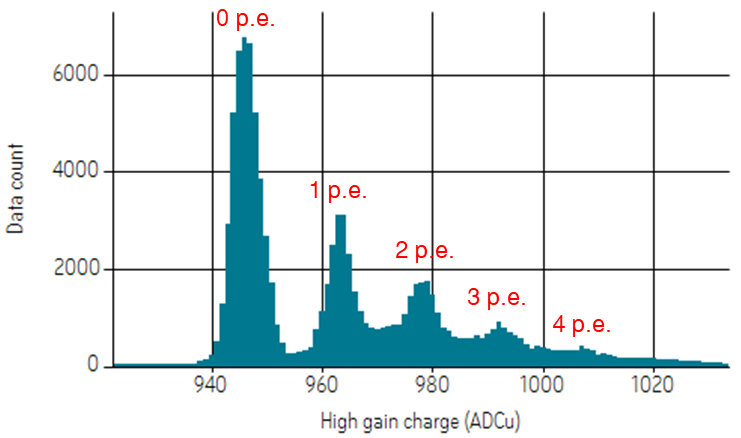}
\caption{Typical charge spectrum obtained with the S13360-3050CS MPPC using the Citiroc 1A evaluation system.}
\label{fig:ChargeDistribution2}
\end{figure}

The 32-channel Citiroc 1A front-end evaluation system from Weeroc was used to read out the MPPC and obtain staircase plots (Fig. \ref{fig:Staircase}) and charge spectra (Fig. \ref{fig:ChargeDistribution2}).

A staircase plot depicts the number of counts per second as a function of threshold. As pulse heights are quantised, the center of each plateau gives the number of \textit{n} photoelectron (p.e.) and above events and can be considered the n - $\frac{1}{2}$ p.e. thresholds where n = 1, 2, 3, and so on. The maximum counting rate of the Citiroc 1A is 20 MHz.

A charge spectrum is a histogram depicting the number of events for each charge integrated over an integration period. This charge is proportional to pulse height. The first peak, or the pedestal, represents the number of 0 p.e. events. Each successive peak after the pedestal contains the number of \textit{n} p.e. events where n = 1, 2, 3, and so on. 

\section{Results}
\subsection{Dark Count Rate}

\begin{figure}[H]
\centering
\includegraphics[width=.6\textwidth]{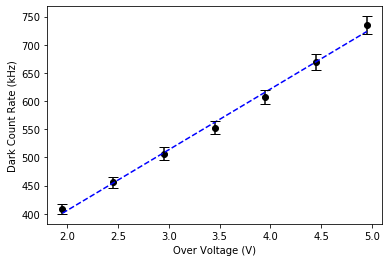}
\caption{Dark count rate as a function of over voltage at 26\textdegree C. The line of best fit is shown by the blue dotted line.} 
\label{fig:DarkCount}
\end{figure}

We measured the average dark count rate as a function of over voltage at 26\textdegree C over three trials. Results can be seen in Figure \ref{fig:DarkCount}. We measured an average dark count rate of 5.07x10\textsuperscript{5} cps at a 2.95V over voltage. The dark count rate increased linearly with over voltage at a rate of approximately 1.08x10\textsuperscript{5} cps per bias volt.


\subsection{Crosstalk Probability}

\begin{figure}[H]
\centering
\includegraphics[width=.6\textwidth]{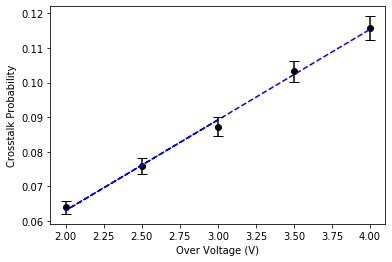}
\caption{Crosstalk probability as a function of over voltage. The line of best fit is shown by the blue dotted line.} \label{fig:Crosstalk}
\end{figure}

We measured the crosstalk probability as a function of over voltage by dividing the frequency of 2 p.e. and above events by the frequency of 1 p.e. and above events in the absence of light. Results can be seen in Fig. \ref{fig:Crosstalk}. We measured a crosstalk probability of 8.7$\%$ at a 2.95V over voltage. The crosstalk probability increased linearly with over voltage at a rate of 5.2$\%$ per bias volt.


\subsection{Photon Detection Efficiency}

We obtained two sets of PDE measurements at 365 nm, 400 nm, 525 nm, 660 nm, 720 nm, 810 nm, and 900 nm using charge integration and pulse counting methods. The charge integration method is described in section \ref{sec:chargeintegration} and the pulse counting method is described in section \ref{sec:pulsecounting}. Results and a comparison of the two methods are presented in section \ref{sec:PDEresults}.

\newpage

\subsubsection{Charge Integration} \label{sec:chargeintegration}


The number of photons incident on the MPPC can be described using the Poisson distribution

\begin{equation}
    P(n;\mu) = \frac{e^{-\mu} \mu^n}{n!}
    \label{eqn:Poisson1}
\end{equation}
where $P(n;\mu)$ is the probability of n events and $\mu$ is the average number of events in an interval. As the pedestal contains the number of 0 p.e. events, it is the only peak that is independent of the effects of crosstalk and afterpulsing and can be used to obtain PDE measurements independent of crosstalk and afterpulsing.   Using \newline n = 0 in Equation \ref{eqn:Poisson1}, we obtain:
\begin{equation}
     P(0;\mu) = \frac{e^{-\mu} \mu^0}{0!} = e^{-\mu}
\label{eqn:Poission}
\end{equation}
Taking the natural logarithm of both sides, we obtain
\begin{equation}
    \mu = -ln(P(0;\mu)) = -ln(\frac{N_{pedestal}}{N_{total}})
\end{equation}
where $N_{pedestal}$ is the number of events in the pedestal and $N_{total}$ is the total number of events taken during the measurement. $N_{total}$ is a user defined variable than can be configured using the Citiroc 1A interface, and $N_{pedestal}$ is found by fitting and integrating the pedestal seen in Fig. \ref{fig:ChargeDistribution2} with a Gaussian of the form
\begin{equation}
    f(x) = \frac{A_{amplitude}}{\sqrt{2\pi\sigma^{2}}}e^{-\frac{(x-\mu)^2}{2\sigma^2}}
\end{equation}
where $A_{amplitude}$ is the amplitude of the Gaussian, $\sigma$ is the standard deviation, and $\mu$ is the mean of the distribution.

After obtaining $\mu_{light}$ for when the MPPC is illuminated and $\mu_{dark}$ for when the MPPC is not illuminated, the number of detected photons per pulse can found by subtracting $\mu_{dark}$ from $\mu_{light}$. As the PDE is defined as number of detected photons divided the number of photons incident on the photosensitive surface of the detector, we use the equation 
\begin{equation}
    PDE = \frac{(\mu_{light} - \mu_{dark})f_{pulse}}{N_{incident}} 
    =\frac{(n_{photons}-n_{dark})f_{pulse}hc}{P_{PM}\lambda} \frac{A_{PM}}{A_{MPPC}}
\end{equation}
where $f_{pulse}$ is the frequency of pulses, $P_{PM}$ is the power measured by the power meter, \textit{h} is Planck's constant, \textit{c} is the speed of light, $A_{MPPC}$ is the photosensitive area of the MPPC, and $A_{PM}$ is the photosensitive area of the power meter. The power detected by the power meter is converted into photons per second using the Einstein-Planck relation, scaled using the ratio between photosensitive surface areas of the power meter and MPPC, and divided by the pulse frequency to find the number of incident photons per pulse.

\subsubsection{Pulse Counting} \label{sec:pulsecounting}

We measured PDE with the pulse counting method using the formula

\begin{equation}
    PDE = \frac{(N_{illuminated} - N_{dark})hc}{\lambda_{LED}P_{PM}}\frac{A_{MPPC}}{A_{PM}}
\end{equation}

where N\textsubscript{illuminated} is the number of counts per second at the 0.5 p.e. threshold when the MPPC is illuminated, N\textsubscript{dark} is the number of dark counts per second at the 0.5 p.e. threshold when the MPPC is not illuminated, $\lambda$\textsubscript{LED} is the wavelength of incident light, P\textsubscript{PM} is the power measured by the power meter at the location of the MPPC, A\textsubscript{MPPC} is the photosensitive area of the MPPC, A\textsubscript{PM} is the photosensitive area of the power meter, h is Planck's constant, and c is the speed of light. Here, we assume each pulse corresponds to only one photon and divide the difference between the number of pulses and the number of dark counts by the number of incident photons. 

\subsubsection{Results} \label{sec:PDEresults}

\begin{figure}[H]
\centering
\includegraphics[width=.6\textwidth]{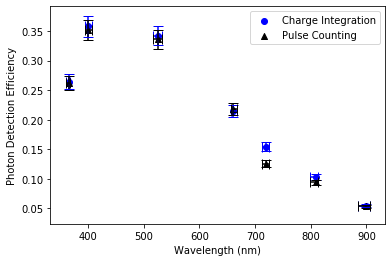}
\caption{PDE as a function of wavelength at  365 nm, 400 nm, 525 nm, 660 nm, 720 nm, 810 nm, and 900 nm using the charge integration and pulse counting methods. Error bars include both statistical and systematic uncertainties.}
\label{fig:PDEComparison}
\end{figure}

\begin{figure}[H]
\centering
\includegraphics[width=.6\textwidth]{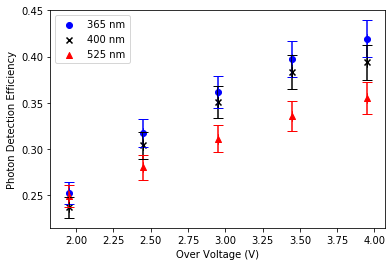}
\caption{PDE as a function of over voltage and wavelength for 365 nm, 400 nm, and 525 nm using the pulse counting method. Error bars include both statistical and systematic uncertainties.}
\label{fig:PDEOvervoltage}
\end{figure}

The PDE as a function of wavelength using the charge integration and pulse counting methods can be seen in Fig. \ref{fig:PDEComparison}, and the PDE as a function of bias voltage for 365 nm, 400 nm, and 525 nm using the pulse counting method can be seen in Fig. \ref{fig:PDEOvervoltage}. We illuminated the MPPC with light pulsed at 800 kHz for the charge integration method and approximately 18 pW of continuous light for the pulse counting method. At 400 nm, we observe a PDE of 36$\%$ using the charge integration method and 35$\%$ using the pulse counting method. On average, we see that the PDE values obtained using the charge integration method are greater than the PDE values obtained using the pulse counting method. This is expected as the pulse counting method cannot distinguish 1 p.e. pulses from 2 p.e. and higher pulses and is blind to multi-photon events. As a result, the pulse counting method is more accurate at low-photon fluxes where the number of 2 p.e. and higher events are negligible. With the exception of 720 nm, both sets of PDE agree to within 1$\%$ PDE. 

\subsection{Linearity and Saturation}

\begin{figure}[H]
\centering
\includegraphics[width=.6\textwidth]{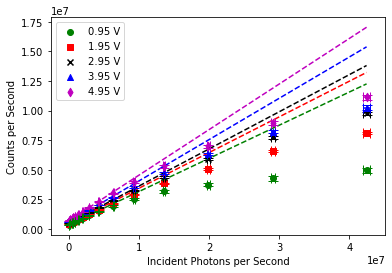}
\caption{Output as a function of incident photon flux and over voltage. Dashed lines correspond to the lines of best fit for data points of their corresponding color in the linear regime below 1.4x10\textsuperscript{6} photons per second.} \label{fig:Linearity}
\end{figure}

We first illuminated the MPPC at five over voltages using a 660 nm LED with incident photon fluxes ranging from 7x10\textsuperscript{5} photons per second to 4.2x10\textsuperscript{7} photons per second  (Fig. \ref{fig:Linearity}). We used continuous light to simulate the response of the MPPC to a continuous sky background. We observed the linear range to increase with over voltage. At a 2.95V over voltage, the output deviates from the line of best fit fitted in the linear region below 1.4x10\textsuperscript{6} photons per second by 10$\%$ at approximately 1.8x10\textsuperscript{7} photons per second.



\begin{figure}[H]
\centering
\includegraphics[width=.6\textwidth]{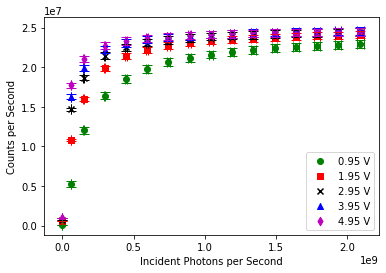}
\caption{Output as a function of incident photon flux and over voltage near and at saturation.}
\label{fig:Saturation}
\end{figure}

We then increased incident photon flux until we reached saturation (Fig. \ref{fig:Saturation}). We find that the saturation threshold decreases with increasing over voltage. This is expected as increasing the over voltage increases dark count rate and PDE of the MPPC while both the maximum number of photons the MPPC can accept, which is a function of the number of pixels, and the maximum 20 MHz counting rate of the Citiroc 1A remain constant. At a 2.95V over voltage, we observe saturation at approximately 5x10\textsuperscript{8} photons per second. 

\section{Conclusion}

In this study, we characterized the dark count rate, crosstalk probability, PDE, linearity, and saturation of the S13360-3050CS MPPC and compared two methods of obtaining PDE. We found an average dark count rate of 5.07x10\textsuperscript{5} cps at 26\textdegree C and a 2.95V over voltage, which is near Hamamatsu's measurement of 500 kcps at 25\textdegree C and a 3V over voltage \cite{datasheet}. To reduce the number of false alarms, we would like to reduce the dark count rate to a negligible level. As the dark count rate of the S13360-3050CS MPPC is expected to halve with each 5\textdegree C drop in temperature\cite{Nepomuk2016}, reducing the operating temperature of the S13360-3050CS MPPC to -40\textdegree C would decrease the dark count rate to approximately 60 cps. Decreasing the operating temperature to lower the dark count rate would be a more feasible method than decreasing the bias voltage as decreasing the bias voltage would also decrease the PDE.  We also measured a crosstalk probability of 8.7$\%$ at a 2.95V over voltage which increased linearly by a rate of 5.2$\%$ per bias volt. While this value is greater than Hamamatsu's measurement of 3$\%$ at a 3V over voltage\cite{datasheet}, we note that the method used in this study does not eliminate the effects of afterpulsing. Future work will involve characterizing both the dark count rate of the S13360-3050CS MPPC and the effectiveness of a coincidence scheme at lowering the number of false alarms caused by dark counts and crosstalk as a functions of operating temperature.

We measured PDE at 365nm, 400 nm, 525 nm, 660 nm, 720 nm, 810 nm, and 900 nm using charge integration and pulse counting methods. At 400 nm, we found a PDE of 36$\%$ using the charge integration method and 35$\%$ using the pulse counting method. Compared to the typical PDE graph provided by Hamamatsu\cite{datasheet}, all values using the charge integration method agree to within statistical and systematic errors except for wavelengths 660 nm and above where we find higher PDE values to within 2$\%$ PDE. On average, we find that the charge integration method produces PDE values greater than PDE values produced by the pulse counting method. This is because unlike the charge integration method the pulse counting method does not distinguish between 1 p.e. and 2 p.e. or higher events and does not eliminate the effects of crosstalk and afterpulsing. As on average the two methods produce PDE values that agree to within 1$\%$ PDE and the charge integration method requires more calculation than the pulse counting method, we conclude that the photon counting method would be suitable method to quickly obtain preliminary PDE measurements at low-photon fluxes.


At 660 nm and a 2.95V overvoltage, the SiPM and Citiroc 1A remain linear until approximately 1.8x10\textsuperscript{7} photons per second and saturate at 5x10\textsuperscript{8} photons per second. The linear range is two orders of magnitude greater than the anticipated sky background of approximately 1.8x10\textsuperscript{5} counts per second per channel and so the MPPC would operate in the linear regime and not be saturated by the sky background at the Assy-Turgen observatory. As these measurements were influenced by the maximum counting rate of the Citiroc 1A, we expect the linear range and saturation of the MPPC itself to be greater than the values obtained here. 


 
Future work to improve the experimental setup will involve incorporating an integrating sphere to improve the uniformity of our light source and a thermoelectric cooler to characterize the dark count rate and PDE as functions of operating temperature. After characterizing other MPPC to compare and select an optimal detector for the UFA telescope, we will begin to characterize larger arrays for camera development and coincidence testing.

The results presented in this paper suggest that the S13360-3050CS MPPC is a feasible detector to use for the UFA telescope at the Assy-Turgen Observatory. These results can be extended to other observatories with similar sky backgrounds requiring a high PDE, low noise single-photon counting detector and searching for fast astrophysical transients.


\acknowledgments 
 
This work was supported by the Hong Kong University of Science and Technology and the Regents' and Chancellor's Research Fellowship from the University of California, Berkeley.
\bibliography{report} 

\begin{thebibliography}{10}

\bibitem{Horowitz2001}
Horowitz, P., Coldwell, C.~M., Howard, A.~B., Latham, D.~W., Stefanik, R.,
  Wolff, J., and Zajac, J.~M., ``{Targeted and all-sky search for nanosecond
  optical pulses at Harvard-Smithsonian},'' in [{\em The Search for
  Extraterrestrial Intelligence (SETI) in the Optical Spectrum
  III}{\nolinebreak\hspace{0.1em}]},  {\em Proc. SPIE}~{\bf 4273},  119--127
  (Aug. 2001).

\bibitem{Eikenberry_1997}
Eikenberry, S.~S., Fazio, G.~G., Ransom, S.~M., Middleditch, J., Kristian, J.,
  and Pennypacker, C.~R., ``{High Time Resolution Infrared Observations of the
  Crab Nebula Pulsar and the Pulsar Emission Mechanism},'' {\em The
  Astrophysical Journal}~{\bf 477},  465--474 (Mar. 1997).

\bibitem{Leung:2018}
Leung, C., Hu, B., Harris, S., Brown, A., Nguyen, H., and Gallicchio, J.,
  ``{Testing the Weak Equivalence Principle using Optical and Near-Infrared
  Crab Pulses},'' {\em The Astrophysical Journal}~{\bf 861},  66 (July 2018).

\bibitem{Li2019}
Li, S., Maire, J., Cosens, M., and Wright, S., ``Detector characterization of a
  near-infrared discrete avalanche photodiode 5x5 array for astrophysical
  observations,'' in [{\em Infrared Technology and Applications
  XLV}{\nolinebreak\hspace{0.1em}]},  {\em Proc. SPIE}~{\bf 11002},  110022G
  (May 2019).

\bibitem{Wright2018}
Wright, S., Horowitz, P., Maire, J., A.~Chaim-Weismann, S., Cosens, M.,
  D.~Drake, F., W.~Howard, A., Marcy, G., P.~V.~Siemion, A., P.~S.~Stone, R.,
  R.~Treffers, R., Uttamchandani, A., and Werthimer, D., ``{Panoramic optical
  and near-infrared SETI instrument: overall specifications and science
  program},'' in [{\em Ground-based and Airborne Instrumentation for Astronomy
  VII}{\nolinebreak\hspace{0.1em}]},  {\em Proc. SPIE}~{\bf 10702},  107025I
  (July 2018).

\bibitem{Rando2015}
Rando, R. et~al., ``{Silicon Photomultiplier Research and Development Studies
  for the Large Size Telescope of the Cherenkov Telescope Array},'' {\em
  PoS}~{\bf ICRC2015},  940 (2016).

\bibitem{Li2019b}
Li, S., Smoot, G.~F., Grossan, B., Lau, A. W.~K., Bekbalanova, M., Shafiee, M.,
  and Stezelberger, T., ``{Program objectives and specifications for the
  Ultra-Fast Astronomy observatory},'' {\em Proc. SPIE}  (2019).
\newblock Manuscript submitted for publication.

\bibitem{Grossan2019}
Grossan, B., Kumar, P., and Smoot, G., ``{The Emission Mechanism of Gamma-ray
  Bursts: Identification via Optical-IR Slope Measurements},'' {\em Journal of
  High Energy Astrophysics}  (2019).
\newblock Manuscript submitted for publication.

\bibitem{ECL}
Grossan, B., ``{Energetic Cosmos Laboratory: Ultra Fast Astronomy}.''
  \url{http://ecl.nu.edu.kz/ultra-fast-astronomy/} (2017).

\bibitem{ECLConf}
Smoot, G., Grossan, B., and Linder, E., ``{ECL Overview}.''
  \url{http://ecl.nu.edu.kz/ultra-fast-astronomy/} (2019).
\newblock Exploring the Energetic Universe 2019 Conference.

\bibitem{Geng2015}
{Geng}, J. and {Huang}, Y., ``{Fast Radio Bursts: Collisions between Neutron
  Stars and Asteroids/Comets},'' {\em The Astrophysical Journal}~{\bf 809},  24
  (Aug. 2015).

\bibitem{Kashiyama2013}
Kashiyama, K., Ioka, K., and Mészáros, P., ``Cosmological fast radio bursts
  from binary white dwarf mergers,'' {\em The Astrophysical Journal}~{\bf 776}
  (July 2013).

\bibitem{Cordes2016}
Cordes, J.~M. and Wasserman, I., ``{Supergiant {P}ulses from {E}xtragalactic
  {N}eutron {S}tars},'' {\em Monthly Notices of the Royal Astronomical
  Society}~{\bf 457},  232--257 (Jan. 2016).

\bibitem{datasheet}
{Hamamtsu Photonics K.K.}, ``{S13360 Series MPPC Datasheet}.''
  \url{https://www.hamamatsu.com/resources/pdf/ssd/s13360_series_kapd1052e.pdf}
  (2016).

\bibitem{Nepomuk2016}
Nepomuk~Otte, A., Garcia, D., Nguyen, T., and Purushotham, D.,
  ``Characterization of three high efficiency and blue sensitive silicon
  photomultipliers,'' {\em Nuclear Instruments and Methods in Physics Research
  Section A: Accelerators, Spectrometers, Detectors and Associated
  Equipment}~{\bf 846} (June 2016).

\end{thebibliography}
\bibliographystyle{spiebib} 

\end{document}